\documentclass{svproc}
\usepackage{subcaption}
\usepackage{fixltx2e}
\usepackage{multirow}
\usepackage{booktabs} % For formal tables
\usepackage{array}
\usepackage{amsmath}
\usepackage{enumitem}
\usepackage[flushleft]{threeparttable} 
\usepackage[framemethod=TikZ]{mdframed}
\usepackage{balance}

\newcommand{\PreserveBackslash}[1]{\let\temp=\\#1\let\\=\temp}
\newcolumntype{C}[1]{>{\PreserveBackslash\centering}p{#1}}
\newcolumntype{R}[1]{>{\PreserveBackslash\raggedleft}p{#1}}
\newcolumntype{L}[1]{>{\PreserveBackslash\raggedright}p{#1}}

\makeatletter
  \newcommand\tablescript{\@setfontsize\tablescript{8.5pt}{6}}
\makeatother

\usepackage{hyperref}

\mdfsetup {outerlinewidth=1pt,
innertopmargin=2pt,
innerbottommargin=2pt,
innerleftmargin=3pt,
innerrightmargin=3pt,
outerlinecolor=gray!15,
middlelinewidth=1pt,
middlelinecolor=white,
backgroundcolor=gray!15,
roundcorner=1pt,
skipabove=.4\baselineskip,
skipbelow=-2pt}

\usepackage{xspace}

\newcommand{\ie}{\textit{i.e.,}\xspace}
\newcommand{\eg}{\textit{e.g.,}\xspace}

\newcommand{\etal}{\textit{et al.}\xspace}

\newcommand{\tdd}{TDD\xspace}
\newcommand{\yw}{YW\xspace}
\newcommand{\bsk}{BSK\xspace}
\newcommand{\mra}{MRA\xspace}

\newcommand{\str}{\texttt{STR}\xspace}
\newcommand{\pls}{PLS\xspace}
\newcommand{\ars}{ARS\xspace}
\newcommand{\dom}{DOM\xspace}
\newcommand{\lik}{LIK\xspace}

\newcommand{\apppls}{\texttt{APP\textsubscript{\pls}}\xspace}
\newcommand{\appars}{\texttt{APP\textsubscript{\ars}}\xspace}
\newcommand{\appdom}{\texttt{APP\textsubscript{\dom}}\xspace}
\newcommand{\applik}{\texttt{APP\textsubscript{\lik}}\xspace}

\newcommand{\imppls}{\texttt{IMP\textsubscript{\pls}}\xspace}
\newcommand{\impars}{\texttt{IMP\textsubscript{\ars}}\xspace}
\newcommand{\impdom}{\texttt{IMP\textsubscript{\dom}}\xspace}
\newcommand{\implik}{\texttt{IMP\textsubscript{\lik}}\xspace}

\newcommand{\tespls}{\texttt{TES\textsubscript{\pls}}\xspace}
\newcommand{\tesars}{\texttt{TES\textsubscript{\ars}}\xspace}
\newcommand{\tesdom}{\texttt{TES\textsubscript{\dom}}\xspace}
\newcommand{\teslik}{\texttt{TES\textsubscript{\lik}}\xspace}

\begin{document}
\mainmatter              % start of a contribution
\title{Results from a Replicated Experiment on the Affective Reactions of Novice Developers when Applying Test-Driven Development}
\titlerunning{Results from a Replicated Experiment on the Affective Reactions of Novice...}  % abbreviated title (for running head)
%                                     also used for the TOC unless
%                                     \toctitle is used
%
\author{Simone Romano\inst{1} \and Giuseppe	Scanniello\inst{2} \and
Maria Teresa Baldassarre\inst{1} \and Davide Fucci\inst{3} \and Danilo Caivano\inst{1}}
\authorrunning{Simone Romano et al.} % abbreviated author list (for running head)
%
%%%% list of authors for the TOC (use if author list has to be modified)
%\tocauthor{Ivar Ekeland, Roger Temam, Jeffrey Dean, David Grove, Craig Chambers, Kim B. Bruce, and Elisa Bertino}
%
\institute{University of Bari, Bari, Italy,\\
\email{\{simone.romano,mariateresa.baldassarre,danilo.caivano\}@uniba.it},\\ 
\and
University of Basilicata, Potenza, Italy,\\
\email{giuseppe.scanniello@unibas.it} \and
Blekinge Institute of Technology, Karlskrona, Sweden\\
\email{davide.fucci@bth.se}}

\maketitle              % typeset the title of the contribution

\begin{abstract}
Test-Driven Development (\tdd) is an incremental approach to software development. Despite it is claimed to improve both quality of software and developers' productivity, the research on the claimed effects of \tdd has so far shown inconclusive results. Some researchers have ascribed these inconclusive results to the negative affective states that \tdd would provoke. A previous (baseline) experiment has, therefore, studied the affective reactions of (novice) developers---\ie 29  third-year undergraduates in Computer Science (CS)---when practicing \tdd to implement software. To validate the results of the baseline experiment, we conducted a replicated experiment that studies the affective reactions of novice developers when applying \tdd to develop software. Developers in the treatment group carried out a development task using \tdd, while those in the control group used a non-\tdd approach.  To measure the affective reactions of developers, we used the Self-Assessment Manikin instrument complemented with a liking dimension. The most important differences between the baseline and replicated experiments are: \textit{(i)}~the kind of novice developers involved in the experiments---third-year vs. second-year undergraduates in CS from two different universities; and \textit{(ii)}~their number---29 vs. 59. The results of the replicated experiment do not show any difference in the affective reactions of novice developers. Instead, the results of the baseline experiment suggest that developers seem to like \tdd less as compared to a non-\tdd approach and that developers following \tdd seem to like implementing code less than the other developers, while testing code seems to make them less~happy.

%We present the results of a replicated experiment that studies the affective reactions of novice developers when applying \tdd  (Test-Driven Development) to develop software. Developers in the treatment group carried out a development task using \tdd, while those in the control group used a non-\tdd approach.  To measure the affective reactions of developers, we used the Self-Assessment Manikin instrument complemented with a liking dimension. The most important differences between the baseline and replicated experiments are: \textit{(i)}~the kind of novice developers involved in the experiments---third-year vs. second-year undergraduates in Computer Science from two different universities; and \textit{(ii)}~their number---29 vs. 59. The results of the replicated experiment do not show any difference in the affective reactions of novice developers. Instead, the results of the baseline experiment suggest that developers seem to like \tdd less as compared to a non-\tdd approach and that developers following \tdd seem to like implementing code less than the other developers, while testing code seems to make them less~happy. Summarizing, the findings of the baseline experiment are not confirmed by the replicated one.  

\end{abstract}

\section{Introduction}
Test-Driven Development (\tdd) is an incremental approach to software development in which unit tests are written before production code~\cite{Beck:2003}. In particular, \tdd promotes short cycles composed of three phases to incrementally implement the functionality of a software:

\begin{description}
\item[Red Phase.] Write a unit test for a small chunk of functionalities not yet implemented and watch the test fail; 
\item[Green Phase.] Implement that chunk of functionalities as quickly as possible and watch all unit tests pass;
\item[Refactor Phase.] Refactor the code and watch all unit tests pass.
\end{description}

Advocates of \tdd claim that this development approach allows improving the (internal and external) quality of software  as well as developers' productivity~\cite{Erdogmus:2010}. However, research on the claimed effects of \tdd, gathered in secondary studies, has so far shown inconclusive results (\eg~\cite{Karac:2018}). Such inconclusive results might relate to the negative affective states that developers would experience when practicing \tdd (\eg ~\cite{Erdogmus:2010}). For example, frustration due to spending a large amount of time in writing unit tests that fail, rather than immediately focusing on the implementation of functionality. Nevertheless, only Romano \etal~\cite{Romano:2020} has studied through a controlled experiment the affective reactions of developers when applying \tdd to implement software. In particular, they recruited 29 novice developers who were asked to carry out a development task by using either \tdd or a non-\tdd approach. At the end of the development task, the researchers gathered the affective reactions to the development approach, as well as to implementing and testing code. To this end, Romano \etal used Self-Assessment Manikin (SAM)~\cite{Bradley:1994}---a lightweight, but powerful self-assessment instrument for measuring affective reactions to a stimulus in terms of the pleasure, arousal, and dominance dimensions---complemented with the liking dimension~\cite{Koelstra:2012}. The results highlight differences in the affective reactions of novice developers to the development approach, as well as to implementing and testing code. In particular, novice developers seem to like \tdd less as compared to a non-\tdd approach. Moreover, novice developers following \tdd seem to like implementing code less than those developers following a non-\tdd approach, while testing code seems to make \tdd developers less~happy. 

The Software Engineering (SE)  community has shown a growing interest in replications of empirical studies (\eg replicated experiments) and recognized the key role that replications play in the construction of knowledge~\cite{DaSilva:2014}. To validate the results of the experiment by Romano \etal~\cite{Romano:2020} (also called baseline experiment from here on), we conducted a replicated experiment with 59 novice developers. In the replication, we investigated the same constructs as the baseline experiment, but in a different site and with participants sampled from a different population---\ie 59 second-year vs. 29 third-year undergraduates in Computer Science (CS) from two different universities. 

\textbf{Paper Structure.} In Section~\ref{sec:background}, we report background information and related work. The baseline experiment is summarized in Section~\ref{sec:baseline}. The replication is outlined in Section~\ref{sec:replication}. The results of our replication are presented and discussed in Section~\ref{sec:results} and Section~\ref{sec:discussion}, respectively.
We discuss the threats to validity of our replication in Section~\ref{sec:threats}. Final remarks conclude the~paper.

\section{Background and Related Work}\label{sec:background}

\begin{figure}[!bt]
\centering
  \includegraphics[width=0.55\linewidth]{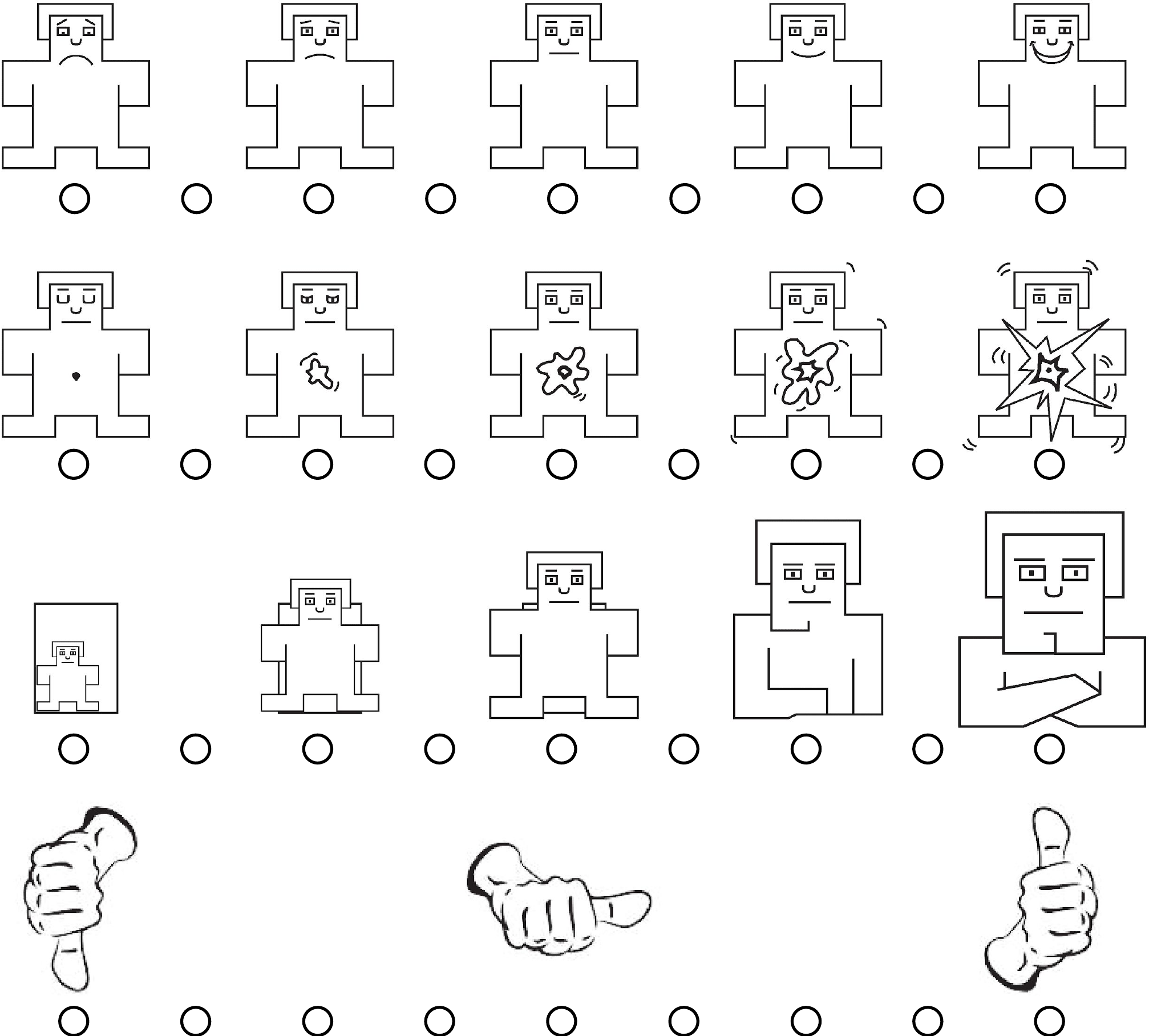}
  \caption{From  top  down, the graphical representations of the pleasure, arousal, dominance, and liking dimensions.  This figure has been taken from~\cite{Romano:2020}.}
  \label{fig:sam}
  %\vspace{-0.5cm}
\end{figure}

According to the PAD (Pleasure-Arousal-Dominance) model---a psychological model to describe and measure affective states---, people's affective states can be characterized through three dimensions: pleasure, arousal, and dominance~\cite{Russell:1977}. The pleasure dimension varies from unpleasant (\eg unhappy/sad) to pleasant (\eg happy/joyful), the arousal one ranges from inactive (\eg bored/calm) to active (\eg excited/stimulated), and finally, the dominance dimension varies from ``without control'' to ``in control of everything''~\cite{Koelstra:2012}. To measure a person's affective reaction to a stimulus in terms of the pleasure, arousal, and dominance dimensions, Bradley and Lang~\cite{Bradley:1994} proposed a pictorial self-assessment instrument they named SAM. This instrument represents each dimension graphically with a rating scale placed just below the graphical representation of each dimension so that a person can self-assess her affective reaction in terms of that dimension (see Figure~\ref{fig:sam}). For instance, SAM pictures the pleasure dimension through manikins varying from an unhappy manikin to a happy one; thus the nine-point rating scale, placed just below the graphical representation of the pleasure dimension, allows a person to self-assess, from one to nine,  that dimension of her affective reaction. Recently, Koelstra \etal~\cite{Koelstra:2012} have complemented SAM with the liking dimension ranging from dislike---pictured through a thumb down---to like---pictured through a thumb up (see Figure~\ref{fig:sam}).

Both Human-Computer Interaction (HCI) and affective computing research fields have utilized SAM in their empirical studies (\eg ~\cite{Herbon:2005,Koelstra:2012}). Later, the SE research field has used SAM as well. For example, Graziotin \etal~\cite{Graziotin:2013} conducted an observational study with eight developers who performed development tasks on individual projects. Every ten minutes, the participants self-assessed both their affective state, by using SAM, and their productivity. The results show that pleasure and dominance are positively correlated with productivity.

A few SE studies have investigated the affective states of developers through controlled experiments (\eg~\cite{Khan:2011,Romano:2019:MTAP}). Besides the study by Romano \etal~\cite{Romano:2020}, which we summarize in the next section, no controlled experiment has been conducted to investigate the affective reactions of developers while practicing~\tdd.

\section{Baseline Experiment}\label{sec:baseline}

In this section, we summarize the baseline experiment by Romano \etal~\cite{Romano:2020} by taking into account the guidelines for reporting replications in SE~\cite{Carver:2014}.

\subsection{Research Questions}
The baseline experiment aimed to answer the following Research Question~(RQ): 
\begin{description}
\item[RQ1.] Is there a difference in the affective reactions of novice developers to a development approach (\ie \tdd vs. a non-\tdd~approach)?
\end{description}
The aim of RQ1 was to understand the affective reactions that \tdd raises on novice developers in terms of pleasure, arousal, dominance, and liking. To deepen such an investigation, two further RQs were formulated and studied:  
\begin{description}
\item[RQ2.] Is there a difference in the affective reactions of novice developers to the implementation phase when comparing \tdd to a non-\tdd approach?
\item[RQ3.] Is there a difference in the affective reactions of novice developers to the testing phase when comparing \tdd to a non-\tdd approach? 
\end{description} 
The aim of RQ2 and RQ3 was to understand the effect of \tdd on the affective reactions of novice developers---in terms of the pleasure, arousal, dominance, and liking dimensions---with respect implementing and testing code, respectively.

\subsection{Participants and Artifacts}
The participants in the baseline experiment were 29 third-year undergraduates in CS at the University of Basilicata (Italy). According to previous work (\eg~\cite{Host:2000}), Romano \etal considered undergraduates in CS as a proxy of novice developers. The participants were taking the SE course when they voluntarily accepted to take part in the experiment. Once the students accepted to participate, they were asked to fill in a pre-questionnaire  (\eg to collect information on their experience on unit testing). Based on the data gathered through this questionnaire, the participants had experience in both C and Java programming. 
No participant had experience with \tdd at the beginning of the SE course. 

The baseline experiment used two experimental objects---\ie Bowling Score Keeper (BSK) and Mars Rover API (MRA). Each participant dealt with either BSK or MRA. The participants, who received BSK, were asked to develop an API for calculating the score of a bowling game, while those who received MRA had to develop an API for moving a rover on a planet. In both cases, they had to code in Java and write unit tests by using JUnit. 
%The participants  did not have to develop a graphic user interface. 
At the beginning of the experimental session, any participant was provided with: \textit{(i)}~a problem statement regarding the assigned experimental object; \textit{(ii)}~the user stories to be implemented (\ie 13 user stories for BSK and 11 user stories for MRA); \textit{(iii)}~a template project for the Eclipse IDE containing the expected API and an example JUnit test class; and \textit{(iv)}~for each user story an acceptance test suite to simulate customers' acceptance of that story. Both BSK and MRA had been previously used as experimental objects in empirical studies on \tdd and could be fulfilled in a three-hour experimental session (\eg~\cite{Fucci:2018:ESEM,Fucci:2016}). 

To gather the affective reactions of the participants, Romano \etal exploited SAM~\cite{Bradley:1994} complemented with the liking dimension~\cite{Koelstra:2012}. SAM allows measuring people's affective reactions to a stimulus over nine-point rating scales in terms of pleasure, arousal, dominance, and liking (see Section~\ref{sec:background}). 

\subsection{Variables and Hypotheses}
The baseline experiment compared the affective reactions of two different groups of novice developers, namely \textit{treatment} and \textit{control}. The treatment group consisted of participants who were asked to use \tdd to carry out a development task, while the control group consisted of participants who were unaware of \tdd and had to perform a development task by using a non-\tdd approach named \yw (Your Way development)---\ie the approach they would normally utilize to develop~\cite{Fucci:2018:ESEM}. Therefore, the main Independent Variable (IV), or main factor, manipulated in the baseline experiment was \textbf{Approach}, which assumed two values: \tdd or \yw. Within each group, some participants dealt with \bsk, while others dealt with \mra. Thus, there was a second IV, namely \textbf{Object}, which had \bsk or \mra as the~value.

To measure the pleasure, arousal, dominance, and liking dimensions with respect to the development approach (\ie to answer RQ1), Romano \etal used the following four ordinal Dependent Variables (DVs): \texttt{\apppls}, \texttt{\appars}, \texttt{\appdom}, and \texttt{\applik}. These variables assumed integer values in between one and nine since each dimension could be assessed through a nine-point rating scale (see Section~\ref{sec:background}). Similarly, they measured pleasure, arousal, dominance, and liking with respect to the implementation and testing phases (\ie to answer RQ2 and RQ3) through the following four ordinal DVs each: \texttt{\imppls}, \texttt{\impars}, \texttt{\impdom}, \texttt{\implik}, \texttt{\tespls}, \texttt{\tesars}, \texttt{\tesdom}, and \texttt{\teslik}. 

To answer the RQs, the following parameterized null hypothesis was tested:
\begin{description}
\item[H0\textsubscript{{DV}.}] There is no effect of Approach on DV $\in \{\texttt{\apppls}, \texttt{\appars}, \texttt{\appdom}, \texttt{\applik}, \\  \texttt{\imppls}, \texttt{\impars}, \texttt{\impdom}, \texttt{\implik}, \texttt{\tespls}, \texttt{\tesars}, \texttt{\tesdom}, \texttt{\teslik}\}$.
\end{description}

\subsection{Design and Execution}\label{sec:design:baseline}
The design of the baseline experiment was 2*2 factorial~\cite{Wohlin:2012}. Such a kind of between-subjects design has two factors (\ie two IVs) having two levels each. The two factors were Approach and Object. Each participant in the baseline experiment was randomly assigned to one development approach and to one experimental object---\ie no participant used both development approaches or dealt with both experimental objects. In particular, 15 participants were assigned to \tdd---7 with \bsk and 8 with \mra---, while 14 participants were assigned to \yw---7 with \bsk and 7 with \mra.

Before the experiment took place, the participants had undergone a training period. In the first part of the training period, all participants attended face-to-face lessons on unit testing, JUnit, Test-Last development (TL), and Incremental Test-Last development (ITL). They also practiced unit testing with JUnit in a laboratory session. In the second part of the training, the participants in the treatment group learned \tdd and practiced it through two laboratory sessions and three homework assignments. 
The participants in the control group did not learn \tdd, rather they practiced TL and ITL through two laboratory sessions and three homework assignments. Regardless of the experimental group, the assignments were the same.
The researcher conducted the experiment in a single three-hour laboratory session at the University of Basilicata where, based on the experimental groups, the participants carried out the development task---\ie they tackled \mra or \bsk---by using \tdd or \yw. At the end of the development task, the participants were asked to self-assess their affective reactions to the used development approach through SAM~\cite{Bradley:1994} complemented with the liking dimension~\cite{Koelstra:2012}. Similarly, they self-assessed their affective reactions to implementing and testing code, respectively. 

\subsection{Data Analysis and Results}
Romano \etal analyzed the effects of Approach, Object, and their interaction (\ie Approach:Object) by using ANOVA Type Statistic (ATS)~\cite{Brunner:1997}, a non-parametric version of ANOVA recommended in the HCI research field to analyze rating-scale data in factorial designs~\cite{Kaptein:2010} (like the case of the baseline experiment). In particular, for each DV, the following ATS~model was built: $DV \sim Approach + Object + Approach:Object$.
To judge whether an effect was statistically significant, the $\alpha$ value was fixed  (as customary) at 0.05. That is, an effect was deemed significant if the corresponding p-value was less~than~$\alpha$. To quantify the magnitude of the effect of Approach, in case it was significant, Romano \etal used Cliff's $\delta$ effect size~\cite{Cliff:1996}. The size of an effect is deemed: \textit{negligible}, if $|\delta|<$ 0.147; \textit{small}, if $0.147\le|\delta|<$ 0.33; \textit{medium}, if $0.33\le|\delta|<$ 0.474; or \textit{large}, otherwise~\cite{Romano:2006}.

In Table~\ref{tab:baseline:ats}, we report the ATS results of the baseline experiment. These results show a significant effect of Approach on \applik (p-value=0.0024), namely there is a significant difference between \tdd and \yw with respect to \applik. This allowed rejecting $H0_{\applik}$. The difference in the \applik values was in favor of \yw and large ($\delta$=0.6048).\footnote{The descriptive statistics were used to determine if the difference was in favor of \tdd or \yw.} Accordingly, Romano \etal concluded that developers using TDD seem to like their development approach less than those using a non-TDD approach (\ie answer to~RQ1). 
Table~\ref{tab:baseline:ats} also shows two further significant effects, one for \implik (p-value=0.0396) and one for \tespls (p-value=0.0178) so allowing rejecting $H0_{\implik}$ and $H0_{\tespls}$, respectively. Both effects were in favor of \yw. The effect size was medium ($\delta$=0.4286) for \implik, while large for \tespls ($\delta$=0.5). Based on these results, Romano \etal concluded that: developers using TDD seem to like the implementation phase less than those using a non-TDD approach (\ie answer to RQ2); and the testing phase seems to make developers using TDD less happy as compared to those using a non-TDD approach (\ie answer to RQ3). As for the effects of Object and Approach:Object, they were in no case significant---\ie neither the experimental object nor the interaction with the development approach seems to influence the affective reactions of novice~developers.

\begin{table}[t]
\centering
\caption{Results, from statistical inference, of the baseline experiment.}\label{tab:baseline:ats}
\begin{threeparttable}
\tablescript
\begin{tabular}{@{}llllll@{}}\toprule
DV                    & \multicolumn{3}{l}{IV} & Cliff's $\delta$ & Outcome for $H0_{DV}$\\
  & Approach & Object & Approach:Object &  &  \\ \midrule
\apppls & 0.1615 & 0.7721 & 0.8998 & - & $H0_{\apppls}$ not rejected \\
\appars & 0.2774 & 0.7794 & 0.1816 & - & $H0_{\appars}$ not rejected \\
\appdom & 0.2796 & 0.8569 & 0.4296 & - & $H0_{\appdom}$ not rejected \\
\applik & 0.0024\tnote{$\ast$} & 0.165 & 0.6368 & 0.6048 (large) & $H0_{\applik}$ rejected in favor of \yw  \\ \addlinespace

\imppls & 0.2008 & 0.6663 & 0.9793 & - & $H0_{\imppls}$ not rejected \\
\impars & 0.6799 & 0.6881 & 0.5752 & - & $H0_{\impars}$ not rejected \\
\impdom & 0.3449 & 0.5614 & 0.4672 & - & $H0_{\impdom}$ not rejected \\
\implik & 0.0396\tnote{$\ast$} & 0.1862 & 0.2703  & 0.4286 (medium) & $H0_{\implik}$ rejected in favor of \yw  \\ \addlinespace

\tespls & 0.0178\tnote{$\ast$} & 0.65 & 0.7652 & 0.5 (large) & $H0_{\imppls}$ rejected in favor of \yw \\
\tesars & 0.4147 & 0.4765 & 0.3406 & - & $H0_{\tesars}$ not rejected \\
\tesdom & 0.6341 & 0.2564 & 0.4738 & - & $H0_{\tesdom}$ not rejected \\
\teslik & 0.0504 & 0.1194 & 0.0547 & - &  $H0_{\teslik}$ not rejected \\ 
\bottomrule
\end{tabular}
\begin{tablenotes}
\item[$\ast$] P-value indicating a significant effect.
\end{tablenotes}
\end{threeparttable}
%\vspace{-0.5cm}
\end{table}

\textbf{Further Analysis and Results.} To better contextualize the baseline experiment, Romano \etal also assessed participants' development performance. To this end, they used a \textit{time-fixed} strategy~\cite{Bergersen:2014}. In particular, they defined an additional DV, named \str, which was computed as follows: \textit{(i)}~count the number of user stories each participant implemented within the fixed time frame (\ie three hours); then \textit{(ii)}~normalize the  number of implemented user stories in $[0,100]$---this is because the total number of user stories of MRA was different to that of BSK (\ie 11 vs. 13). It is ease to grasp that the higher the \str value is, the better the development performance of a given participant is. Romano \etal analyzed the effects of Approach, Object, and Approach:Object on \str by using ATS because the normality assumption to apply ANOVA~\cite{Wohlin:2012} was not met. The results of ATS did not indicate a significant effect of Approach (p-value = 0.4765) on \str, namely the development approach seems not to influence the participants' development performance. The effects of Object (p-value = 0.2596), and Approach:Object (p-value = 0.0604) on \str were not significant. 

\begin{table}[t]
\centering
\caption{Summary of baseline and replicated experiments.}\label{tab:summary}
\tablescript
\begin{tabular}{@{}lp{4.6cm}p{4.6cm}@{}}\toprule
Characteristic & Baseline experiment & Replication \\ \midrule
Participant type & III-year undergraduates in CS taking the SE course at the University of Basilicata & II-year undergraduates in CS taking the SE course at the University of Bari\\ \addlinespace
Participant number & 29 & 59 \\ \addlinespace
Site & University of Basilicata & University of Bari\\ \addlinespace
RQs & RQ1, RQ2, RQ3 & RQ1, RQ2, RQ3 \\ \addlinespace
Experimental objects & BSK, MRA  & BSK, MRA \\ \addlinespace
Experimental groups & TDD, YW & TDD, YW \\ \addlinespace
Environment & Java, Eclipse, JUnit & Java, Eclipse, JUnit \\ \addlinespace 
Design & 2*2 factorial & 2*2 factorial \\ \addlinespace
\multirow{2}{2.6cm}{Assignment to groups and objects} & 15 participants assigned to \tdd (7~\bsk, 8~\mra), 14 participants assigned to \yw (7~\bsk, 7~\mra) & 28 participants assigned to \tdd (14~\bsk, 14~\mra), 31 participants assigned to \yw (16~\bsk, 15~\mra)\\ \addlinespace
IV & Approach, Object & Approach, Object \\ \addlinespace
DV & \texttt{\apppls}, \texttt{\appars}, \texttt{\appdom}, \texttt{\applik}, \texttt{\imppls}, \texttt{\impars}, \texttt{\impdom}, \texttt{\implik}, \texttt{\tespls}, \texttt{\tesars}, \texttt{\tesdom}, \texttt{\teslik} & \texttt{\apppls}, \texttt{\appars}, \texttt{\appdom}, \texttt{\applik}, \texttt{\imppls}, \texttt{\impars}, \texttt{\impdom}, \texttt{\implik}, \texttt{\tespls}, \texttt{\tesars}, \texttt{\tesdom}, \texttt{\teslik} \\ \addlinespace
Null hypotheses & ${H0_{DV}}$ & ${H0_{DV}}$ \\ \addlinespace
\multirow{2}{2.6cm}{Statistical inference method} & ATS to analyze the effects of Approach, Object, and Approach:Object & ATS to analyze the effects of Approach, Object, and Approach:Object \\ 
\bottomrule
\end{tabular}
%\vspace{-0.5cm}
\end{table}

\section{Replicated Experiment}\label{sec:replication}

We conducted a replicated experiment to determine whether the results from the baseline experiment are still valid  in a different site and with a larger number of participants sampled from a different population. Despite these differences, we designed and executed the replicated experiment as similarly as possible to the baseline experiment to determine, in case of inconsistent results with the baseline experiment, which factors could have caused those results. To this end, we used the replication package of the baseline experiment, which is available on the web\footnote{\href{https://doi.org/10.6084/m9.figshare.9778019.v1}{https://doi.org/10.6084/m9.figshare.9778019.v1}} and  includes experimental objects, analysis scripts, and raw data. 

As shown in Table~\ref{tab:summary}, the replicated experiment shares most of the characteristics of the baseline one. Therefore, in the following of this section, we limit ourselves to describe the replicated experiment in terms of participants, and design and execution. This is to say that RQs, artifacts, variables, hypotheses, and data analysis of the replication are the same as the baseline experiment; therefore, such information can be found in Section~\ref{sec:baseline}.

\subsection{Participants}
The participants in the replication were 59 second-year undergraduates in CS at the University of Bari who were taking the SE course. Participation was on a voluntary basis (\ie we did not pay the students for their participation). To encourage students to participate in the replication, we rewarded the participants with two bonus points in the final mark of the SE course (as had been done in the baseline experiment). The two bonus points were given regardless of the performance of the participants in the replication.
Similarly to the baseline experiment, the participants were asked to fill in a pre-questionnaire. Based on the participants' answers, they had passed the exams of the Basic and Advanced Programming courses and had experience with C and Java programming. The participants were not knowledgeable in \tdd. 

\subsection{Design and Execution}\label{sec:design:replication}
Based on the 2*2 factorial design used in the baseline experiment, the participants in the replication were randomly assigned to the experimental groups and objects: 28 participants were assigned to \tdd---14 with \bsk and 14 with \mra---; while 31 participants were assigned to \yw---16 with \bsk and 15 \mra.

All the participants in the replication attended face-to-face lessons on unit testing, JUnit, TL, and ITL. They also practiced unit testing with JUnit in a laboratory session. Later, the participants in the treatment group learned \tdd and practiced it through two laboratory sessions and two homework assignments. 
The participants in the control group, who did not learn \tdd, practiced TL and ITL through two laboratory sessions and two homework assignments. The material (\eg homework assignments) used to train the participants was the same as the baseline experiment, although the number of the homework assignments was different between the baseline and replicated experiments---\ie three vs. two. We were forced to give two homework assignments, rather than three, because the students could not carry out a third homework assignment during the training period due to deadlines that other courses requested in the same period. As so, we preferred not overloading students to avoid  threat of dropouts from the experiment. We conducted the experiment in a single three-hour laboratory session in which the participants carried out the development task---\ie they tackled \mra or \bsk---by using \tdd or \yw based on their experimental group. At the end of the development task, the participants self-assessed their affective reactions to the used development approach, as well as to implementing and testing code, by using SAM~\cite{Bradley:1994} complemented with the liking dimension~\cite{Koelstra:2012}.

%\vspace{+0.5cm}
\begin{figure}[t]
   \centering
    \includegraphics[width=\textwidth]{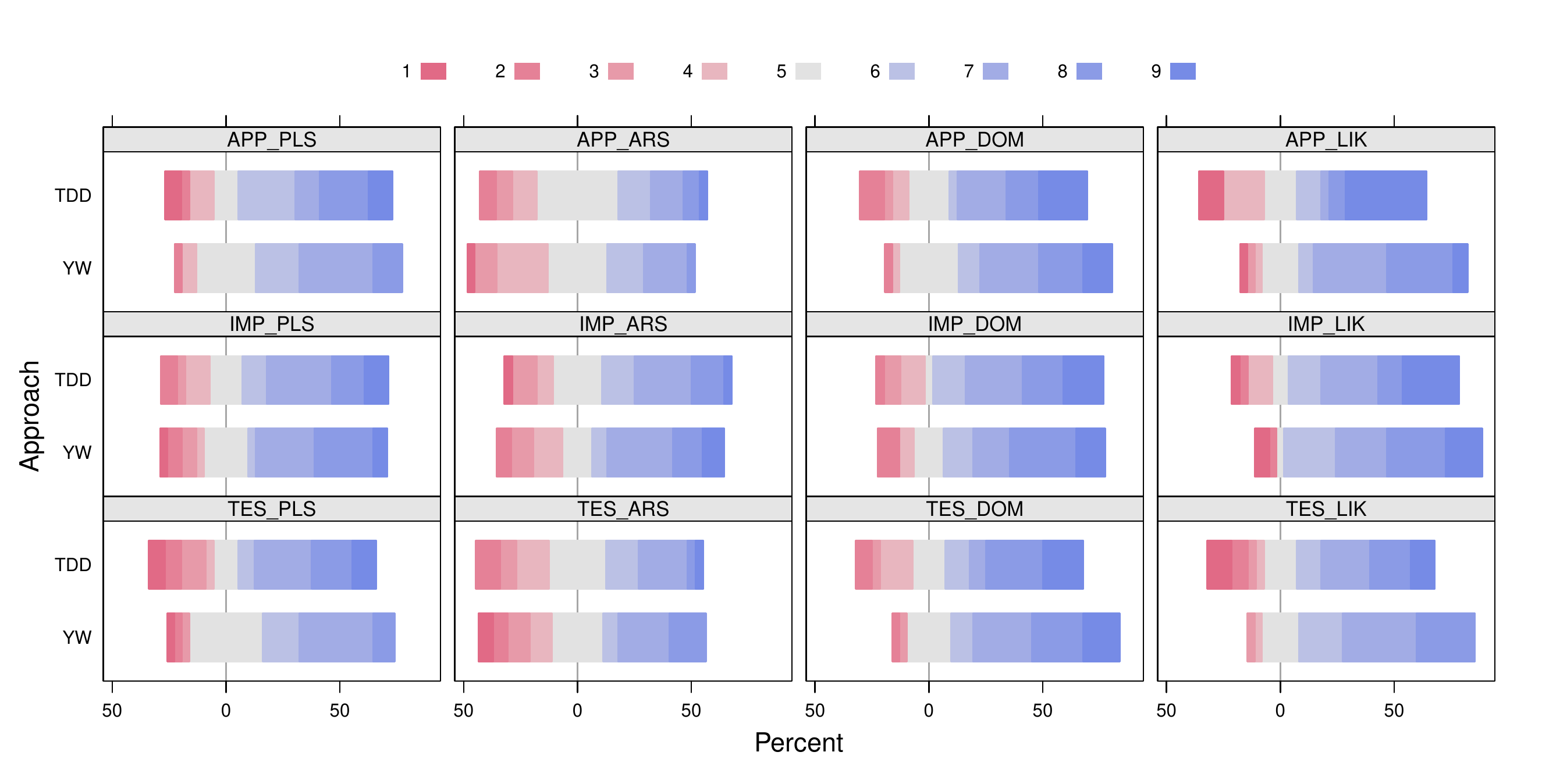}
    \caption{Diverging stacked bar plots summarizing the DV values of the~replication.}\label{fig:plot}
    %\vspace{-0.5cm}
\end{figure}

\section{Results}\label{sec:results}
In Figure~\ref{fig:plot}, we summarize the values of the DVs (of the replicated experiment) by using diverging stacked bar plots. These plots show the frequencies of the DV values grouped by Approach. For each DV, the neutral judgment (\ie five) is displayed in grey; while negative judgments (\ie from one to four) and those positive (\ie from six to nine) are shown in shades of red and blue, respectively. The width of a colored bar (\eg the grey one) is proportional to the frequencies of the corresponding DV value (\eg five in the corresponding DV value for the grey bar).
The interested reader can find the raw data on the web.\footnote{\href{https://doi.org/10.6084/m9.figshare.12085821.v1}{https://doi.org/10.6084/m9.figshare.12085821.v1}} The p-values ATS returned for each DV are reported in Table~\ref{tab:ats}.

\begin{table}[t]
\centering
\tablescript{}
\caption{Results, from statistical inference, of the replication.}\label{tab:ats}
\begin{threeparttable}
\begin{tabular}{@{}lllll@{}}\toprule
DV                    & \multicolumn{3}{l}{IV} &  Outcome for $H0_{DV}$ \\ \cmidrule{2-4}
                  & Approach & Object & Approach:Object & \\ \midrule
\apppls & 0.6937 & 0.0805 & 0.7001 & $H0_{\apppls}$ not rejected  \\
\appars & 0.6421  &  0.9018  & 0.2817 & $H0_{\appars}$ not rejected \\
\appdom & 0.8295  & 0.1376  & 0.5235 & $H0_{\appdom}$ not rejected  \\
\applik & 0.9211 & 0.0324\tnote{$\ast$} & 0.2571 & $H0_{\applik}$ not rejected \\ \addlinespace

\imppls & 0.904 & 0.2849 & 0.4421 &  $H0_{\imppls}$ not rejected \\
\impars & 0.7781 & 0.9646 & 0.3198 & $H0_{\impars}$ not rejected \\
\impdom & 0.9529 & 0.2389 & 0.9411 & $H0_{\impdom}$ not rejected \\
\implik & 0.8048 & 0.1314 & 0.6618 & $H0_{\implik}$ not rejected \\ \addlinespace

\tespls & 0.5722 & 0.3083 & 0.7749 & $H0_{\tespls}$ not rejected \\
\tesars & 0.7446 & 0.2281 & 0.4129 & $H0_{\tesars}$ not rejected \\
\tesdom & 0.509 & 0.1079 & 0.9945 &  $H0_{\tesdom}$ not rejected \\
\teslik & 0.4588 & 0.3457 & 0.1566 & $H0_{\teslik}$ not rejected \\ 
\bottomrule
\end{tabular}
\begin{tablenotes}
\item[$\ast$] P-value indicating a significant effect.
\end{tablenotes}
\end{threeparttable}
%\vspace{-0.5cm}
\end{table}

\textbf{RQ1---Affective Reactions to the Development Approach.} The plots in Figure~\ref{fig:plot} (see the first row) do no show huge differences in the affective reactions to the used development approach, namely \tdd or \yw, in terms of pleasure (\apppls), arousal (\appars), dominance (\appdom), and liking (\applik). However, it seems that \tdd has some negative frequencies more than \yw as far as the dominance and liking dimensions are concerned. 
The results of ATS (see Table~\ref{tab:ats}) indicate that there is no significant effect of Approach on the pleasure, arousal, dominance, and liking dimensions of the participants' affective reactions to the development approach. Accordingly, we cannot reject the corresponding null hypotheses. Finally, we did not find any significant effect of the interaction between Approach and Object, while the effect of Object is significant on the liking dimension (p-value=0.0324). That is, the used experimental object significantly influenced the affective reactions of the participants to the development approach in terms of liking. However, the effect of the experimental object is consistent within both experimental groups as there is no significant~interaction.

\begin{mdframed}
\noindent\textbf{Answer to RQ1.} We observed no significant difference in the affective reactions of novice developers to the used development approach, \ie \tdd~or~\yw.
\end{mdframed}

\textbf{RQ2---Affective Reactions to the Implementation Phase.}
As shown in Figure~\ref{fig:plot}, there is no huge difference between \tdd and \yw regarding pleasure (\imppls), arousal (\impars), dominance (\impdom), and liking (\implik) of the affective reactions to the implementation phase. We can also notice that, as for the liking dimension, \tdd seems to have some negative frequencies more than \yw. The results of ATS (see Table~\ref{tab:ats}) do not show any significant effect of Approach on the four dimensions. Therefore, the corresponding null hypotheses cannot be rejected. The effects of Object and its interaction with Approach are not significant. 

\begin{mdframed}
\noindent
\textbf{Answer to RQ2.} With respect to the implementation phase, the results do not show a significant difference in the affective reactions of novice developers when they use \tdd or \yw.
\end{mdframed}

\textbf{RQ3---Affective Reactions to the Testing~Phase.}
The plots in Figure~\ref{fig:plot} show that the affective reactions of the control group to the testing phase in terms pleasure (\tespls), arousal (\tesars), dominance (\tesdom), and liking (\teslik) are similar to the those of the treatment group. However, except for the arousal dimension, a slight trend in favor of \yw can be observed since there are more negative frequencies for \tdd as compared to \yw. The results in Table~\ref{tab:ats} do not allow rejecting the null hypotheses. Finally, neither the effect of Object nor its interaction with Approach is~significant. 

\begin{mdframed}
\noindent
\textbf{Answer to RQ3.} We did not observe  a significant difference in the affective reactions of novice developers to the testing phase when they use \tdd or~\yw.
\end{mdframed}

\textbf{Further Analysis Results.} We used ATS to analyze \str because the normality assumption of ANOVA was not met (Shapiro-Wilk normality test p-value = 0.001). The results of ATS do not indicate a significant effect of Approach (p-value = 0.448) on \str, while the effect of Object (p-value \textless 0.001) was significant so suggesting that there was a difference in the development performance of the participants when dealing with BSK or MRA. However, the effect of the experimental object is consistent within both experimental groups since the interaction Approach:Object (p-value = 0.566) is not significant.

\section{Discussion}\label{sec:discussion}
Replications that do not draw the same conclusions as the baseline experiment can be viewed as successful, on a par with replications that come to the same conclusions as the baseline experiment~\cite{Shull:2008}. Our replication falls into the former case since the outcomes of the replicated experiment do not fully confirm the outcomes of the baseline one. In particular, the baseline experiment found that participants seem to: \textit{(i)} like \tdd less as compared to \yw; \textit{(ii)} like less implementing code with \tdd; and \textit{(iii)} be less happy when testing code using \tdd. The replication cannot support these findings because we did not observe any significant difference between \tdd and \yw. As for the other investigated constructs (\eg arousal due to the used development approach), the outcomes of the baseline experiment are confirmed by those of the replicated experiment (\ie the statistical conclusions are the same). 

\begin{figure}[t]
\begin{subfigure}{0.5\textwidth}
  \centering
  % include first image
  \includegraphics[width=1\linewidth]{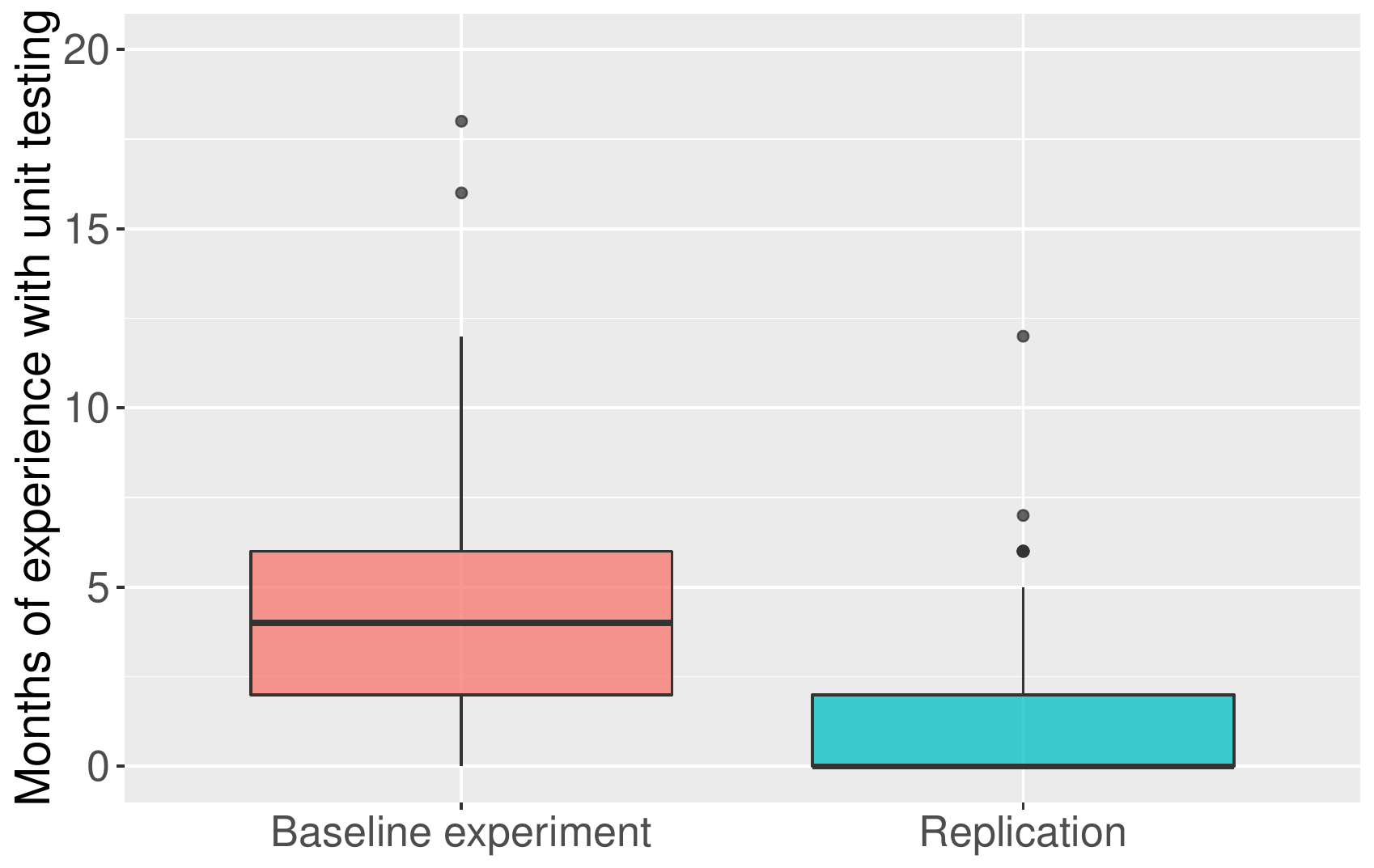}  \caption{}\label{fig:plot:testing}
\end{subfigure}
\begin{subfigure}{.5\textwidth}
  \centering
  % include second image
  \includegraphics[width=1\linewidth]{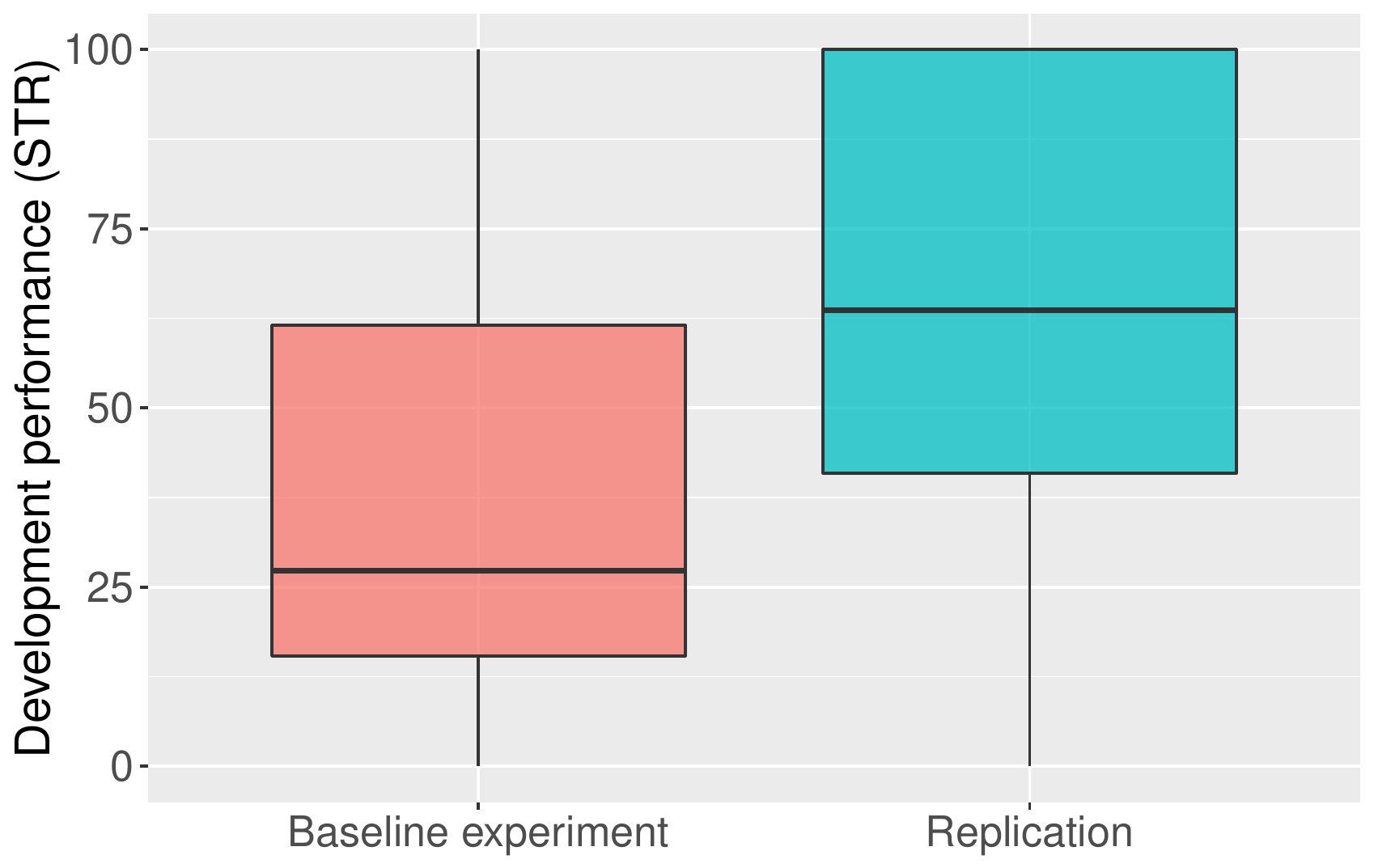}  
  \caption{}
  \label{fig:plot:str}
\end{subfigure}
\caption{Box-plots summarizing (a) months of experience with unit testing (at the beginning of the SE courses) of the participants and (b) development performance of the participants in the baseline and replicated experiments.}
\label{fig:fig}
%\vspace{-0.5cm}
\end{figure}

The question that now arises is why the replication fails to fully support the findings of the baseline one. We speculate that the inconsistent results between the baseline and replicated experiments are due to the type of participants (third-year vs. second-year undergraduates in CS from two different universities), rather than their number (29 vs. 59). Although the number of participants in the baseline experiment was not so high and less than that of the participants in the replication, the magnitude (\ie Cliff's $\delta$ effect size) of the three significant effects~\cite{Caivano2005}, in the baseline experiment, was either medium or large. Such a magnitude makes us quite confident that the inconsistent results between the baseline and replicated experiments are not due to the number of participants. This is why we ascribe them to the type of participants. In particular, the participants in the baseline experiment were more experienced with unit testing than those in the replication, who mostly had no experience (see Figure~\ref{fig:plot:testing}). Since the participants in the baseline experiment did not know \tdd (at the beginning of the SE course in which the experiment was run), they were therefore used to practice unit testing in a \textit{test-last} manner. That is, they were used to write unit tests after they had written production code---in contrast to \tdd, where unit tests are written before producing code. This is to say that the participants in the baseline experiment were probably more conservative and therefore less prone to change the order with which they usually wrote production and testing code. Accordingly, their affective reactions, due to TDD, were more negative. This postulation suggests two possible future research directions: \textit{(i)} replicating the baseline experiment with more experienced developers to ascertain that the greater the experience with unit testing in a test-last manner, the more negative their affective reactions, due to TDD, are; and \textit{(ii)} conducting an observational study with a cohort of developers to investigate if the affective reactions caused by \tdd change over time. The above-mentioned postulation could be of interest to lecturers teaching unit testing. In particular, they could start teaching \tdd as soon as possible to lessen/neutralize the negative affective reactions that \tdd causes; after all, there is empirical evidence showing that, with time, \tdd leads developers to write more unit~tests~\cite{Fucci:2018:ESEM}.

Another characteristic of the participants that varies between the baseline and replicated experiments is the academic year of the CS program in which the participants were enrolled---\ie third year vs. second one. This implies that the participants in the baseline experiment have learned to code in Java a few months before than those in the replication. Nevertheless, the development performance was better in the replication than in the baseline experiment (see Figure~\ref{fig:plot:str}). Therefore, we are quite confident that the academic year did not cause the inconsistent results between the baseline and replicated experiments. On the other hand, we cannot exclude that the worse development performance of the participants in the baseline experiment could have somehow amplified the differences in the affective reactions of the participants who practiced \tdd or \yw. After all, past work (\eg~\cite{Graziotin:2013,Khan:2011}) has found that the affective states of developers are related to their performance in SE tasks, despite it is still unclear the role that \tdd can play in such a relation. To better investigate this point, we suggest researchers to replicate the baseline experiment by introducing a change in the design, namely: allowing any participant to fulfil the development task (\ie no fixed time), rather than giving any participant a fixed time frame to carry the development task. Such a design choice should allow isolating the effect that the development performance could have on the affective reactions of developers.

\section{Threats to Validity}\label{sec:threats}
The replicated experiment inherits most of the threats to validity of the baseline one since, in the replicated experiment, we introduced few changes. We discuss the threats to validity according to the guidelines by Wohlin \etal~\cite{Wohlin:2012}.  

\textbf{Construct validity.} Threats concern the relation between theory and observation~\cite{Wohlin:2012}. We measured each DV once by using a self-assessment instrument (\ie SAM). As so, in case of measurement bias, this might affect the obtained results (threat of \textit{mono-method bias}). Although we did not disclose the research goals of our study to the participants, they might have guessed them and changed their behavior based on their guess (threat of \textit{hypotheses guessing}). To mitigate a threat of \textit{evaluation apprehension}, we informed the participants that they would get two bonus points on the final exam mark regardless their performance in the replication. There might be a threat of \textit{restricted generalizability across constructs}. That is, \tdd might have influenced some non-measured constructs. 

\textbf{Conclusion validity.} Threats concern issues that affect the ability to draw the correct conclusion~\cite{Wohlin:2012}.
We mitigated a threat of \textit{random heterogeneity of participants} through two countermeasures: \textit{(i)} we only involved students taking the SE course allowing us to have a sample of participates with similar background, skills, and experience; \textit{(ii)} the participants underwent a training period to make them as more homogeneous as possible within the groups. A threat of \textit{reliability of treatment implementation} might have occurred. For example, a few participants might have followed \tdd more strictly than others, somehow influencing  their affective reactions. To mitigate this threat, during the experiment, we reminded the participants to use the development approach we  assigned them. 
Although SAM is one of the most reliable instruments for measuring affective reactions~\cite{Morris:2002}, there might be a threat of  \textit{reliability of measures} since the measures gathered by using SAM, as well as the liking scale, are subjective in~nature. 

\textbf{Internal validity.} Threats are influences that can affect the IVs with respect to the causal relationship between treatment and outcome~\cite{Wohlin:2012}. A \textit{selection} threat might have affected our results since the participation in the study was on a voluntary basis and volunteers might be more motivated to carry out a development task than the whole population of developers. Another threat that might have affected our results is \textit{resentful demoralization}, namely participants assigned to a less desirable treatment might not behave as they normally would. 
To mitigate a possible threat of \textit{diffusion or treatments imitations}, we monitored the participants during the execution of the replication and alternated the participants dealing with \bsk to those dealing with \mra.

\textbf{External validity.} Threats to external validity concern the  generalizability of results~\cite{Wohlin:2012}.
In the replication, we involved undergraduates in CS to reduce the heterogeneity among the participants. This implies that generalizing the results to the population of professional developers might lead to a threat of \textit{interaction of selection and treatment}. That is, while we mitigated a threat to conclusion validity like \textit{random heterogeneity of participants}, we could not mitigate a threat to external validity. We prioritized a threat of \textit{random heterogeneity of participants} to better determine, in case of different results between the baseline and replicated experiments, which factors might have caused such inconsistent results. However, it is worth mentioning that: \textit{(i)}~the use of students could be appropriate as suggested in the literature (\eg~\cite{Host:2000,Lemos:2012,Salman:2015}) and \textit{(ii)}~the development performance of the participants in the replication was better than that of the participants in the baseline experiment (see Figure~\ref{fig:plot:str}).
The use of \bsk and \mra as experimental objects might represent a threat of \textit{interaction of setting and treatment} despite they are commonly used as experimental objects in empirical studies on \tdd (\eg ~\cite{Fucci:2018:ESEM,Fucci:2016,Salman:2015}). Moreover, both \bsk and \mra can be fulfilled in a single three-hour laboratory session~\cite{Fucci:2018:ESEM} so allowing better control over the~participants.

\section{Conclusion}
We conducted a replicated experiment on the affective reactions of novice developers when applying \tdd to implement software. With respect to the baseline experiment, we varied the experimental context and number of participants. The results from the replicated experiment do not fully confirm those of the baseline one. We speculate that the kind of  developers can influence the affective reactions due to \tdd. In particular, developers who have experience with unit testing in a test-last manner could have affective reactions, due to \tdd, that are more negative than  developers who have no/little experience with unit testing in a test-last manner. We also speculate that developers' performance in implementing software can influence the affective reactions of developers when applying \tdd.

\bibliographystyle{splncs03}
\bibliography{bibliography}

\end{document}